\documentclass[twocolumn,showpacs,preprintnumbers,amsmath,amssymb]{revtex4}

\usepackage{graphicx}
\usepackage{dcolumn}
\usepackage{bm}
\usepackage{epsfig}
\usepackage{longtable} 

\begin{document}

\preprint{}

\title{
Measurement of the half-life of $^{198}$Au in a non-metal:  High-precision measurement shows no host-material dependence
}

\author{J. R. Goodwin}

\author{N. Nica}

\author{V. E. Iacob}

\author{A. Dibidad}
\altaffiliation {REU summer student from Florida A\&M University, Tallahassee, Florida 32307, USA}

\author{J. C. Hardy}
\email{hardy@comp.tamu.edu}

\affiliation{ Cyclotron Institute, Texas A\&M University, College Station, Texas 77843, USA}
\homepage{http://cyclotron.tamu.edu/}

\date{\today}

\begin{abstract}
We have measured the half-life of the $\beta^-$ decay of $^{198}$Au to be 2.6948(9) d, with the nuclide sited
in an insulating environment.  Comparing this result with the half-life we measured previously with a metallic
environment, we find the half-lives in both environments to be the same within 0.04\%, thus contradicting a
prediction that screening from a ``plasma" of quasi-free electrons in a metal increases the half-life by as
much as 7\%.
 
\end{abstract}


\maketitle

\section{\label{sec:introd} INTRODUCTION}

This experiment was undertaken to investigate if the half life of the $\beta^-$ decay of $^{198}$Au
depends on whether the decaying nucleus is located in a metallic or an insulating environment.  The
``Debye plasma model", which was originally invoked \cite{Ra04} to explain observed cross-section anomalies
in the d(d,p)t reaction, was later applied to radioactive decays by Limata {\it et al.}\,\cite{Li06}.  According
to this model, the conduction electrons present in a metal comprise a sort of plasma, which is referred
to as a Debye plasma.  It has been argued that this plasma changes the phase space available for radioactive decay, and
increases (for $\beta^-$ or electron-capture decay) or decreases (for $\beta^+$ decay) the nuclide's
half-life.  If this model were correct, this change in phase space would occur only in metals, not in
insulators; and would be enhanced if the metal were to be cooled to very low temperatures.

In their subsequent study of the $\beta^-$ decay of $^{198}$Au sited in a pure-gold host material, Spillane
{\it et al.}\,\cite{Sp07} claimed to have observed both these effects, albeit to a lesser extent than the theory
predicted.  The theory predicts that at room temperature the half-life of $^{198}$Au sited in a metal
should be 7\% longer than it is in an insulator, while at 12\,K the difference should increase even further
to 32\%.  The corresponding measured numbers as reported by Spillane {\it et al.}, were 0.4(7) and 4.0(7)\%.  
We repeated their measurement in a metal at two different temperatures and have already reported \cite{Go07} that
any temperature dependence must be less than 0.04\%, two orders of magnitude below the value claimed by
Spillane {\it et al.}  However, we have not yet addressed the possibility that there might be a difference
between a $^{198}$Au source distributed in a metal and one in an insulator.  We do so now by reporting a
measurement of the $^{198}$Au half-life, for which the decaying nuclei were sited in Au$_2$O$_3$.

Both measurements of the half-life of $^{198}$Au in gold metal -- ours \cite{Go07} and that of Spillane
{\it et al.} \cite{Sp07} -- were performed with sources prepared by neutron activation of natural gold, 
$^{197}$Au.  To obtain comparable conditions and statistics for our measurement in a non-metal, we wished to use
neutron activation again and sought a suitable gold compound that is also an insulator.  Although strictly
speaking it is not an insulator, we did identify Au$_2$O$_3$ -- gold (III) oxide -- as a suitable candidate.  It
is considered to be a semiconductor \cite{Sh07} but, with a calculated band gap higher than 0.85 eV, it should
behave like an insulator at room temperature.  In fact, it does: Its room-temperature resistivity has been
measured to be at least five orders of magnitude higher than that of pure gold \cite{Ma00}, undoubtedly
sufficient to ensure the absence of a conduction-electron plasma.  

\section{\label{sec:Appar} Apparatus and set-up}

Gold has two important advantages for precise half-life measurements: it is monoisotopic ($^{197}$Au), so
neutron activation produces only $^{198}$Au; and its decay spectrum is dominated by a single strong $\beta$-delayed
$\gamma$ ray at 412 keV.  No corrections are required for contaminant activities and the peak-to-background ratio
is very high.  Although we used Au$_2$O$_3$ as the material to be activated in this experiment, all other aspects of
the measurement were identical to those of our previous experiment \cite{Go07}, in which we activated pure gold.  We
can thus directly compare the $^{198}$Au half-lives measured at room-temperature with two different host materials, 
one an insulator and the other a conductor.

We used a gold (III) oxide sample obtained from the Alfa Aesar Corporation.  It was in the form of powder with a
purity of 99.99\%.  A 170-mg quantity of this powder was held onto an aluminum disc by adhesive Mylar tape, 56 $\mu$m
thick, and the assembly was activated in a flux of $\sim$10$^{10}$ neutrons/cm$^{2}$s for 10 s at the Texas A\&M Triga
reactor.  The irradiated Au$_2$O$_3$ sample was then fastened on the cold head of a CryoTorr
7 cryopump, precisely as had been done previously for our pure gold measurement \cite{Go07}.  Although we did not cool
the Au$_2$O$_3$ sample to a low temperature in this measurement, for consistency we nevertheless followed the same
procedure as in the previous measurement, including the use of the cryopump as a location for our sample.

A 70\% HPGe detector was placed directly facing the sample on the cryopump axis just outside the pump's cover plate, 
into which a cavity had been bored so that only 3.5 mm of stainless steel remained between the sample and the face of the
detector.  This arrangement was not altered in any way throughout the decay measurement.  Sequential six-hour
$\gamma$-ray spectra were acquired and recorded for a total period of 27 days -- 10 half-lives
of $^{198}$Au.  The detector signals were amplified and sent to an analog-to-digital converter, which was an Ortec
TRUMP$^{\textsc{TM}}$-8k/2k card \cite{Ortec} controlled by Maestro software, which was installed on a PC operating under
Windows-XP.

During the entire period of the measurements, our computer clock was synchronized daily against the signal
broadcast by WWVB, the radio station operated by the U.S. National Institute of Standards and Technology.  The
TRUMP$^{\textsc{TM}}$ card uses the Gedcke-Hale method \cite{Je81} to correct for dead time losses, so by keeping our system's
dead time below about 3\% and recording all our spectra for an identical pre-set live time, we ensured that our results
were nearly independent of dead time losses.  However, to achieve a precision better than 0.1\% a further small
correction is required to account for residual rate-dependent effects such as pulse pile-up.  As described in
Ref.~\cite{Go07}, we have experimentally determined the fractional residual loss for our system to be
5.5(2.5)\,$\times$\,10$^{-4}$ per 1\% increase in dead time.  We applied this correction to the present results as we
also did for the measurement to which this one is being compared: the half-life of $^{198}$Au in gold metal.

\begin{figure}[t]
\epsfig{file=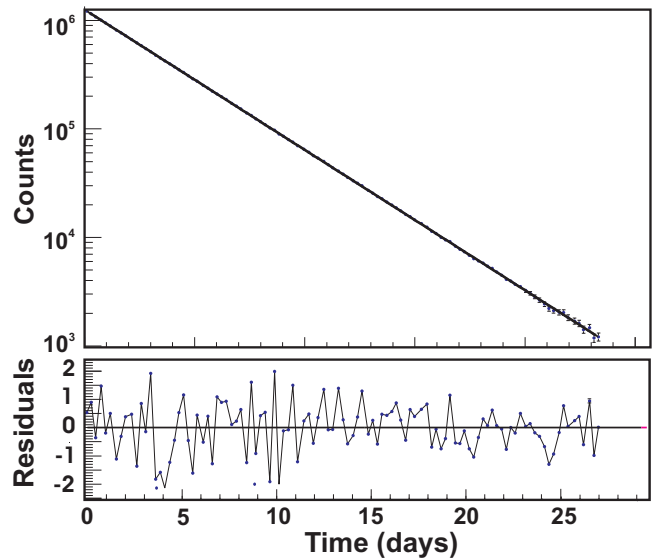,width=8.5cm}
\caption{The decay of $^{198}$Au in gold (III) oxide, at room temperature. Experimental data appear as dots in
the upper portion of the figure; the straight line is the fit to these data.  Normalized residuals are shown at the bottom.}
\label{fig1}
\end{figure}

\section{\label{sec:Res} Analysis and Results}

We analyzed the 412-keV $\gamma$-ray peak in each of the recorded spectra by using the least-square peak-fitting
program GF3 from the RADware series \cite{Ra01}.  Use of this program allowed us to be very specific in determining
the correct background for a peak, and we visually inspected the peak of interest in each spectrum to ensure that
the background was handled satisfactorily.  So far as possible, the same criteria were applied to each of the 107
recorded spectra.  The peak areas thus obtained for the 412-keV peak were then corrected for residual losses as
described in Sec.~\ref{sec:Appar}.  The results are plotted as a function of time in Fig.~\ref{fig1}.

The decay curve was then analyzed by a maximum-likelihood fit with a single exponential.  The code we used, which
is based on ROOT \cite{Br97}, has previously been tested to a precision of 0.01\% with Monte Carlo generated data.
The result of the fit for the gold oxide half-life measurement is shown in Fig.~1, where the fitted decay curve
is compared with the data in the top panel, and the normalized residuals are plotted in the bottom panel.  The $^{198}$Au
half life obtained from this fit (with statistical uncertainty only) is 2.6948(9)~d.  The corresponding normalized
$\chi^2$ is 0.74, which gives a confidence level of 99\%.

The equivalent room-temperature result for the $^{198}$Au half life, as measured in a pure-gold host material,
was reported by us \cite{Go07} to be 2.6949(5)~d.  The difference between these two results is 0.0001(10)~d or
0.004(38)\%.  Both measurements were made under the same conditions and the data from both have been corrected
for residual losses; however, the uncertainty in that correction has not been applied since it is correlated
for the two measurements and does not contribute to the difference between them.

For the present measurement the systematic uncertainty associated with the residual-loss correction is small compared
to the statistical uncertainty, so the total uncertainty is unchanged from the statistical one.  Our final result
is thus 2.6948(9)~d.  This is in excellent agreement with 2.69500(27)~d, the weighted average of all previous
measurements of the gold half-life (see Ref.~\cite{Go07}), most of which were performed at an unrecorded
temperature and in an unspecified host medium.

\section{\label{sec:Conc} Conclusions }

We have measured the half-life of $^{198}$Au in gold (III) oxide at room temperature.  This result obtained with
the decaying nuclei sited in this insulating medium is consistent with a half-life result we published previously
for $^{198}$Au sited in pure gold, a conductor.  We established that the difference between the half-lives
measured in an insulator and in a conductor is less than 0.04\%, with a confidence level of 68\% (one standard
deviation).  This limit is more than two orders of magnitude
lower than the 7\% difference predicted by the ``Debye plasma model" \cite{Sp07}.  Our result, together with
previous measurements of ours \cite{Go07,Go09} and others \cite{Ru08}, effectively refutes all the predictions
of the Debye plasma model as they apply to $\beta^-$, $\beta^+$ and electron-capture decays; and also contradicts
the measurements that initially supported those predictions \cite{Sp07,Li06,Wa06}.

Our concern in undertaking these measurements was for the integrity of precise half-lives measured in the past.  
Since physical conditions were believed to have no influence on half-lives, no care was taken in the past to select a
particular host material or even to specify the temperature at which a measurement was made.  Our main concern
was with the half-lives of superallowed 0$^+$$\rightarrow$0$^+$ $\beta^+$ emitters, which are essential to
fundamental tests of the standard model \cite{Ha09}.  Their precision has typically been quoted to less than
0.05\%, well below the temperature and host-material dependence claimed by the measurements in
Refs.~\cite{Sp07,Li06,Wa06}.  We can now state with confidence that, at the level of 0.05\%, half-lives are
neither affected by temperature changes between 19\,K and 295\,K, nor by the resistivity of the host medium
in which they are located.  There is no need to revisit past measurements of half-lives quoted to high precision.

\begin{acknowledgments}

We thank Prof.~R. Watson for his interest and for helpful discussions.  We also appreciate the assistance of the
staff at the Texas A\&M Nuclear Science Center, where we had our source activated.  This work was supported by the
U.S. Department of Energy under Grant No. DE-FG03-93ER40773 and by the Robert A. Welch Foundation under Grant No.
A-1397.  

\end{acknowledgments}


\begin{thebibliography}{}

\bibitem{Ra04}
F. Raiola, L. Gang, C. Bonomo, G. Gy\"urky, M. Aliotta, H.W. Becker, R. Bonetti, C. Broggini,
P. Corvisiero, A. D'Onofrio, Z. F\"ul\"op, G. Gervino, L. Gialanella, M. Junker, P. Prati, V. Roca, 
C. Rolfs, M. Romano, E. Somorjai, F. Strieder, F. Terrasi, G. Fiorentini, K. Langanke and J. Winter, 
Eur. Phys. J. A {\bf 19}, 283 (2004).

\bibitem{Li06}
B. Limata, F. Raiola, B. Wang, S. Yan, H.W. Becker, A. D'Onofrio, L. Gialanella, V. Roca, C. Rolfs, 
M. Romano, D. Sch\"urmann, F. Strieder and F. Terrasi, Eur. Phys. J. A {\bf 28}, 251 (2006).

\bibitem{Sp07}
T. Spillane, F. Raiola, F. Zeng, H.W. Becker, L. Gialanella, K.U. Kettner, R. Kunze, C. Rolfs, 
M. Romano, D. Sch\"urmann and F. Strieder, Eur. Phys. J. A {\bf 31}, 203 (2007).

\bibitem{Go07}
J. R. Goodwin, V.V. Golovko, V.E. Iacob and J.C. Hardy, Eur. Phys. J. A. {\bf 34}, 271 (2007).

\bibitem{Sh07}
H. Shi, R Asahi and C Stampfl, Phys. Rev. B {\bf 75}, 205125 (2007).

\bibitem{Ma00}
F. Machalett, K. Edinger, J. Melngailis, M. Diegel, K, Steenbeck, E. Steinbeiss, Appl. Phys. A {\bf 71}, 331 (2000).

\bibitem{Ortec}
http://www.ortec-online.com/trump.htm

\bibitem{Je81}
R. Jenkins, R.W. Gould, D. Gedcke, Quantitative X-ray Spectrometry (Marcel Dekker, New York, 1981) pg. 266.

\bibitem{Ra01}
D. Radford, http://radware/phy.ornl.gov/main.html (private communication).

\bibitem{Br97}
R. Brun and F. Rademakers, Nucl. Instrum. Methods A {\bf 389}, 81(1997).

\bibitem{Go09}
J. R. Goodwin, V.V. Golovko, V.E. Iacob and J.C. Hardy, Phys. Rev. C {\bf 80} 045501 (2009).

\bibitem{Ru08}
G. Ruprecht, C. Vockenhuber, L. Buchmann, R. Woods, C. Ruiz, S. Lapi and D. Bemmerer, Phys. Rev. C 77,
065502 (2008) and C 78, 039901(E) (2008).

\bibitem{Wa06}
B. Wang, S. Yan, B. Limata, F. Raiola, M. Aliotta, H.W. Becker, J. Cruz, N. De Cesare, A. D'Onofrio, 
Z. F\"ul\"op, L. Gialanella, G. Gy\"urky, G. Imbriani, A. Jesus, J.P. Ribeiro, V. Roca, D. Rogalla, 
C. Rolfs, M. Romano, D. Sch\"urmann, E. Somorjai, F. Strieder and F. Terrasi, Eur. Phys. J. A 28, 375 (2006).

\bibitem{Ha09}
J.C. Hardy and I.S. Towner, Phys. Rev. C 79, 055502 (2009).



\end{thebibliography}
\end{document}